\documentclass{PoS}
\usepackage{graphicx}
\usepackage{txfonts}

\def\beq#1{\begin{equation}\label{#1}}
\def\eeq{\end{equation}}
\def\beqa#1{\begin{eqnarray}\label{#1}}
\def\eeqa{\end{eqnarray}}
\def\eq#1{eq.~(\ref{#1})}
\def\Eq#1{Eq.~(\ref{#1})}

\def\myfrac#1#2{\left(\frac{#1}{#2}\right)}

\def\mycomment#1{\relax}

\title{Wind Accretion - Observations Vs Theory}

\ShortTitle{Wind Accretion - Observations Vs Theory}

\author{\speaker{N. Shakura}%
        \thanks{Supported by RSF grant 14-12-00146}\\
       Sternberg Astronomical Institute, Moscow M.V.Lomonosov State University\\
       E-mail: \email{nikolai.shakura@gmail.com}}
\author{{K. Postnov}\\
       Sternberg Astronomical Institute, Moscow M.V.Lomonosov State University\\
       E-mail: \email{kpostnov@gmail.com}}

\abstract{The theory of quasi-spherical subsonic accretion onto magnetized 
rotating neutron star is reviewed. Different regimes of quasi-spherical accretion onto a neutron star: supersonic (Bondi)
accretion, which takes place when the captured matter cools down rapidly and falls
supersonically towards the neutron-star 
magnetosphere, and subsonic (settling) accretion which occurs when
the plasma remains hot until it meets the magnetospheric boundary. 
In subsonic accretion, which works at
X-ray luminosities $\lesssim 4\times 10^{36}$~erg~s$^{-1}$, a hot quasi-spherical shell 
must form around the magnetosphere, and the
actual accretion rate onto the neutron star is determined by the ability of the plasma to enter the
magnetosphere due to the Rayleigh-Taylor instability. We show how the dimensionless 
parameters of the theory can be determined from observations of equilibrium X-ray pulsars
(Vela X-1, GX 301-2).  
We also discuss how in the settling accretion theory bright X-ray flares ($\sim 10^{38}-10^{40}$~ergs) observed in supergiant fast X-ray transients (SFXT)  may be produced by 
sporadic capture of magnetized stellar-wind plasma. At sufficiently low accretion rates, 
magnetic reconnection 
can enhance the magnetospheric plasma entry rate, resulting in copious production of X-ray photons,
strong Compton cooling and ultimately in unstable 
accretion of the entire shell. 
A bright flare develops on the free-fall time
scale in the shell, and the typical energy released in an SFXT bright flare corresponds to 
the mass of the shell.
 }

\FullConference{Accretion Processes in Cosmic Sources\\
                 5-10 September 2016\\
                 Saint Petersburg, Russia}

\begin{document}

\section{Introduction: Short history of X-ray astronomy and accretion}
\label{intro}
On June 18, 1962, a serendipitous discovery of the first galactic X-ray source, Sco X-1, was made 
\cite{1962PhRvL...9..439G}. The project was originally aimed at observing X-ray fluorescent emission from the Moon, but instead this discovery heralded the beginning of X-ray astronomy. The fluorescent X-ray emission from the Moon was actually discovered about 20 years later by the ROSAT satellite \cite{1991Natur.349..583S}. Sco X-1 was the brightest 
galactic X-ray source far beyond the Solar system, and later a lot of interesting galactic X-ray sources 
(Cyg X-1, Her X-1, Cen X-3, etc.) were discovered in other rocket experiments. Before the launch of the specialized X-ray satellite UHURU (12 December 1970, \cite{1971ApJ...165L..27G}), the origin of the powerful X-ray emission from galactic sources was unclear. However, as early as in the mid-1960, Yakov Zeldovich 
\cite{1964DAN} and Ed Salpeter \cite{1964ApJ...140..796S} invoked accretion of matter onto moving compact 
objects as powerful source of energy emission. First UHURU results showed that galactic X-ray sources can 
be quasi-persistent (like Cyg X-1 \cite{1971ApJ...166L...1O}) or show periodic pulsations (like Cen X-3 \cite{1971ApJ...167L..67G} and Her X-1 \cite{1972ApJ...174L.143T}). Later it was recognized that disk accretion onto 
a compact star in a binary system is responsible for the observed powerful X-ray emission \cite{1973A&A....24..337S}. In close binary systems, accretion disks are formed during mass transfer 
from the optical star onto compact stellar remnants (neutron stars or black holes) 
through the vicinity of the inner Lagrangian point. In the case of black holes, accretion disks extend down the to the last marginally stable circular orbit ($6GM/c^2$ for a Schwarzschild black hole). 
In the case of magnetized neutron stars, the magnetic field of neutron star starts destroying accretion flow at distances typically about 100-1000 NS radii. The accreting matter enters the NS magnetosphere, gets frozen into the NS magnetic field and is canalized to the NS magnetic polar caps, where most of the accretion power is emitted.  
The disk accretion regime is usually realized 
when the optical star overfills its Roche lobe. If the optical star does not fill its Roche lobe, accretion still 
can be very powerful from the captured stellar wind \cite{1973A&A....24..337S,1973ApJ...179..585D}. Even in this case accretion disk can be formed if the specific angular momentum of captured matter is high enough; if not, accretion flow will 
be quasi-spherical. In this review we will consider only quasi-spherical accretion onto magnetized NSs.

\section{Two regimes of wind accretion}

 Quasi-spherical accretion is 
most likely to occur in high-mass X-ray binaries (HMXB) 
when the optical star of early spectral class (OB) does not fill its Roche lobe, 
but experiences a significant mass loss via stellar wind. We shall discuss the wind accretion regime,
in which a bow shock forms in the stellar wind around the compact star.
The characteristic distance at which the bow shock forms
is about the gravitational capture (Bondi) radius 
\beq{e:R_B}
R_B=2GM/(v_w^2+v_{orb}^2)\,,
\eeq 
where $v_w$ is the wind velocity 
(typically 100-1000 km/s), $v_{orb}$ is the orbital velocity of NS, which is usually much smaller than $v_{w}$, so below we will neglect it. 
The rate of gravitational capture of mass from the wind with density $\rho_w$
near the orbital position of the NS  
is the Bondi-Hoyle-Littleton mass accretion rate: 
\beq{e:dotM_B}
\dot M_B\simeq \rho_w R_B^2 v_w\propto \rho_w v_w^{-3}\,.
\eeq

\subsection{Supersonic (Bondi-Hoyle-Littleton) accretion}  
 
There can be two different cases of quasi-spherical accretion. The classical 
Bondi-Hoyle-Littleton 
accretion takes place when the shocked matter rapidly cools down, 
and the matter freely falls towards the NS magnetosphere 
(see Fig. \ref{f:1}) by forming 
a shock at some distance above the magnetosphere. 
Here the shocked matter cools down (mainly by Compton processes) 
and enters the magnetosphere via Rayleigh-Taylor instability \cite{1976ApJ...207..914A}.
The magnetospheric
boundary is characterized by the Alfv\'en radius $R_A$, which can be
calculated from the balance of the ram pressure of the 
infalling matter and the magnetic field pressure at the 
magnetospheric boundary: $\rho v_{ff}^2(R_A)=B^2/8\pi$. Making use of the mass continuity equation in the shell, $\dot M=4\pi R^2\rho(R)v_{ff}(R)$, and assuming dipole NS magnetic field, 
the standard result \cite{1973ApJ...179..585D} is obtained:
\beq{e:R_Astand}
R_A=\myfrac{\mu^2}{\dot M\sqrt{2GM}}^{2/7}\,.
\eeq
The captured matter from the wind carries a specific angular momentum 
$j_w\sim \omega_BR_B^2$ \cite{1975A&A....39..185I}. 
Depending on the sign of $j_w$ (prograde or retorgrade), the NS can spin-up 
or spin-down. This regime of quasi-spherical accretion occurs in 
bright X-ray pulsars with $L_x>4\times 10^{36}$~erg~s$^{-1}$ \cite{1983ApJ...266..175B,2012MNRAS.420..216S}. 

\subsection{Subsonic (settling) accretion}

\begin{figure*}
\includegraphics[width=7cm]{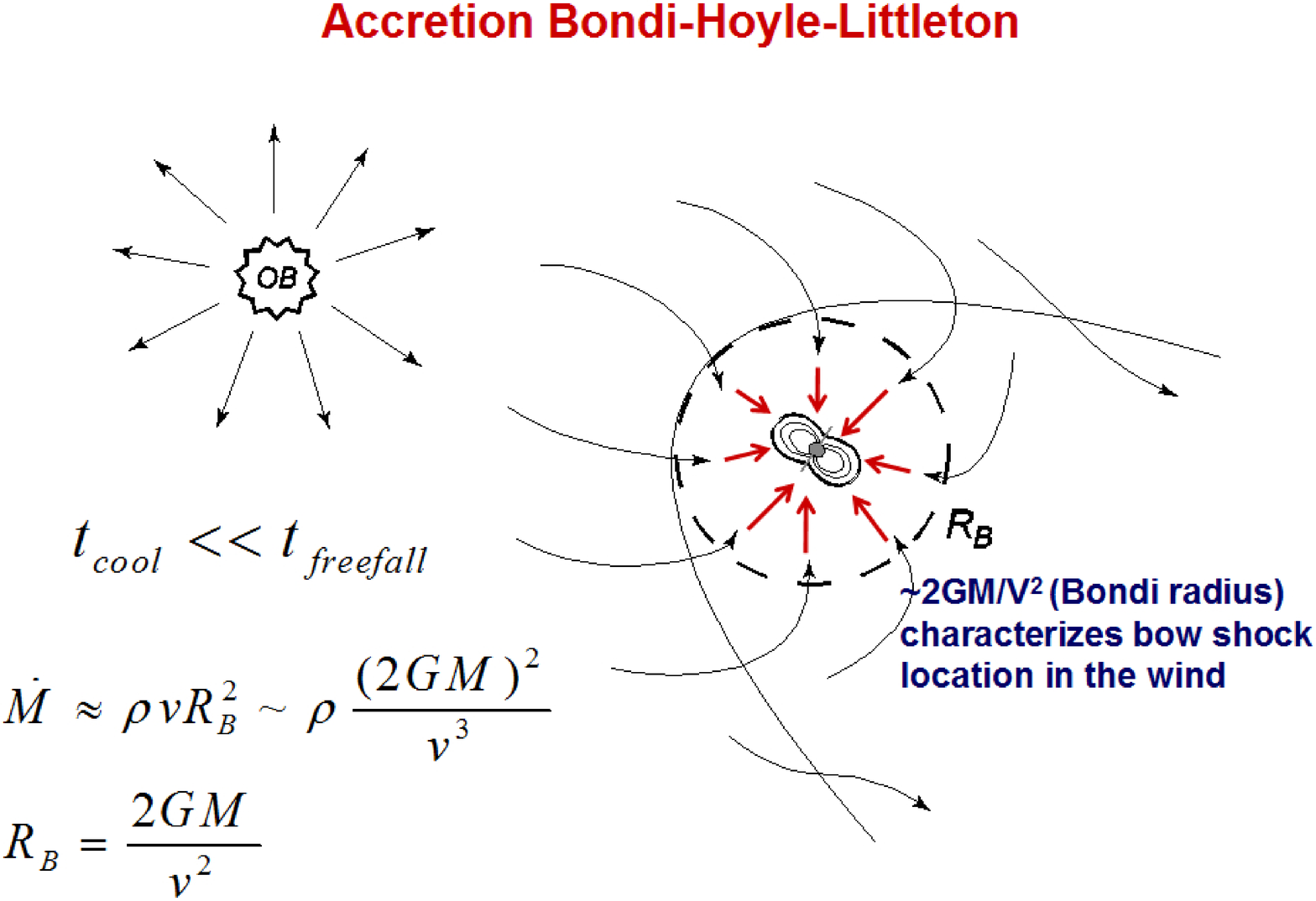}
\includegraphics[width=7cm]{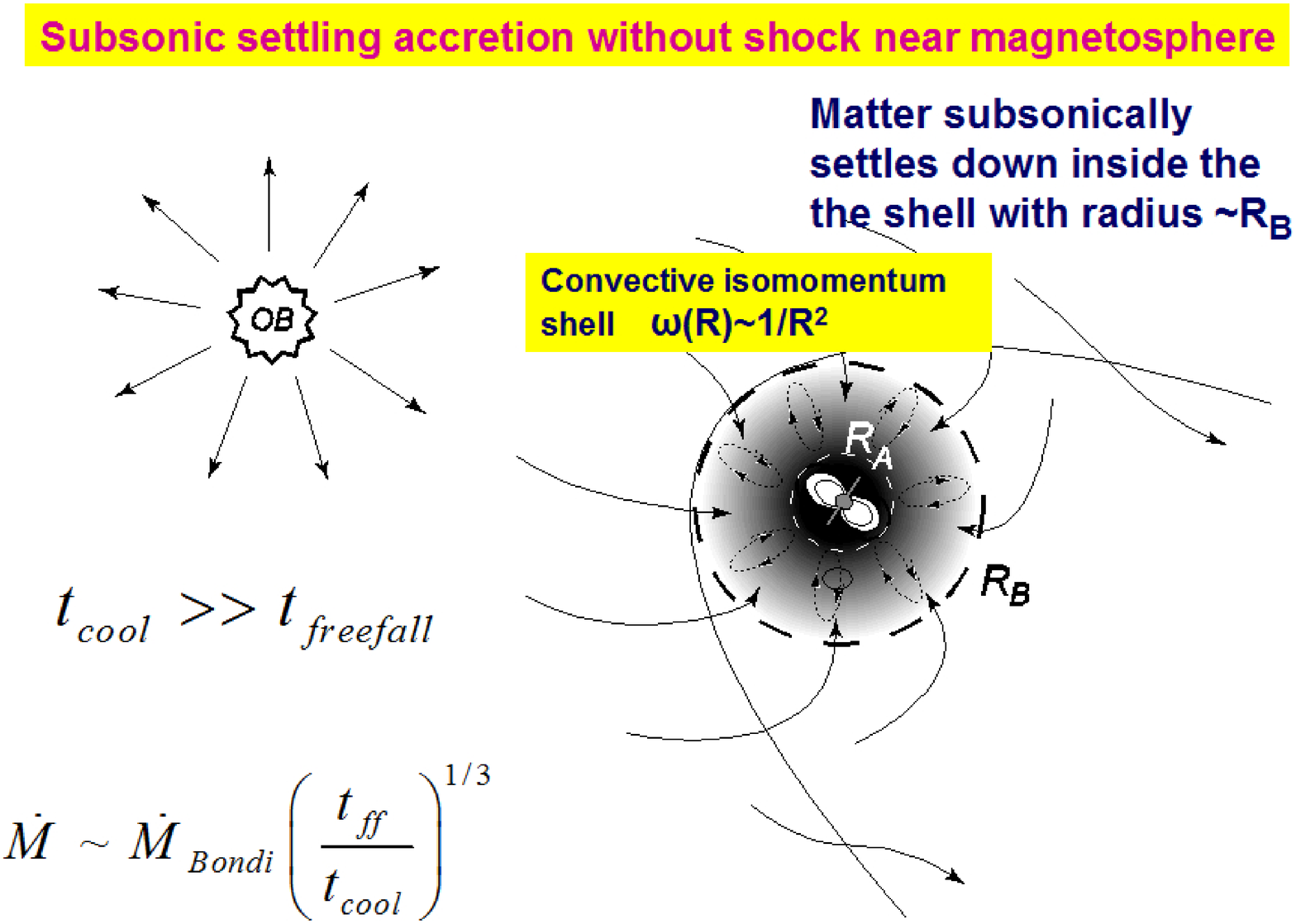}
\parbox[t]{0.47\textwidth}{\caption{Supersonic (Bondi-Hoyle-Littleton) accretion onto magnetized NS}\label{f:1}}
\hfill
\parbox[t]{0.47\textwidth}
{\caption{Subsonic settling accretion onto magnetized NS}\label{f:2}}
\end{figure*}

If the captured wind matter behind the bow shock at $R_B$ remains hot (when 
the plasma cooling time 
is much longer than the free-fall time, $t_{cool}\gg t_{ff}$), 
a hot quasi-static shell forms around the magnetosphere. The subsonic 
(settling) accretion sets in (see Fig. \ref{f:2}).  
In this case, both spin-up or spin-down of the NS 
is possible, even if the sign of $j_w$ is positive (prograde). The shell mediates the
angular momentum transfer from the NS magnetosphere via viscous stresses
due to convection and turbulence. In this regime, the mean radial velocity 
of matter in the shell $u_r$ is smaller than the free-fall velocity $u_{ff}$: 
$u_r=f(u)u_{ff}$, $f(u)<1$, and is determined by the plasma cooling rate 
near the magnetosphere (due to the Compton or radiative cooling): 
\beq{e:f(u)}
f(u)\sim [t_{ff}(R_A)/t_{cool}(R_A)]^{1/3}.
\eeq 
In the settling accretion regime 
the actual mass accretion rate
onto NS can be significantly smaller than the Bondi mass accretion rate, 
\beq{}
\dot M=f(u) \dot M_B\,.
\eeq  
The settling accretion occurs 
at $L_x<4\times 10^{36}$~erg~s$^{-1}$ \cite{2012MNRAS.420..216S}. 

\section{Structure of the shell}

The structure of the shell around NS magnetosphere in the settling accretion regime is 
discussed in detail in \cite{2012MNRAS.420..216S}. To the first approximation, its vertical 
structure along the radius $R$ can be described assuming hydrostatic equilibrium:
\beq{}
-\frac{1}{\rho}\frac{dP}{dR}=\frac{GM}{R^2}
\eeq
with the adiabatic solution for the temperature radial profile
\beq{e:T(R)}
\frac{{\cal R}T}{\mu_m}=\frac{\gamma-1}{\gamma}\frac{GM}{R}\,.
\eeq
For the adiabatic index $\gamma=5/3$ we get the standard result (see also \cite{1981MNRAS.196..209D}):
\beq{e:rho(R)}
\rho(R)=\rho(R_A)\myfrac{R_A}{R}^{3/2}\,,
\eeq
where $R_A$ is the magnetospheric (Alfv'en) radius. 

Unlike the supersonic Bondi regime, in the settling accretion regime the magnetospheric 
boundary is determined by balance between the gas thermal pressure and magnetic pressure yielding \cite{2012MNRAS.420..216S}
\beq{RAdef}
R_A=\left[\frac{4\gamma}{\gamma-1}f(u)K_2\frac{\mu^2}{\dot M\sqrt{2GM}}\right]^{2/7}\,,
\eeq
where the factor $K_2\simeq 7.6$ takes into account the effect of magnetospheric currents \cite{1976ApJ...207..914A}. Clearly, in the settling accretion regime the dependence of $R_A$ on $\dot M$ and
$\mu$ can be different than in the standard formula \Eq{e:R_Astand}, since the factor $f(u)$ depends
differently on $\dot M$ and $\mu$ for different cooling regime. Numerically, $f(u)\sim 0.1-0.5$ depending on the X-ray luminosity. 
The plasma enters the magnetosphere of the slowly rotating neutron star due to 
the Rayleigh-Taylor instability. The boundary between the plasma and the magnetosphere is stable 
at high temperatures $T>T_{cr}$, but becomes unstable at $T<T_{cr}$, and remains in 
a neutral equilibrium at $T=T_{cr}$ \cite{1977ApJ...215..897E}. The critical temperature is:
\beq{Tcr}
{\cal R}T_{cr}=\frac{1}{2}\frac{\cos\chi}{\kappa R_A}\frac{\mu_mGM}{R_A}\,.
\eeq 
Here $\kappa$ is the local curvature of the magnetosphere, $\chi$ is the angle 
the outer normal to the magnetospheric surface makes with the radius-vector at a given point. The effective gravity acceleration can be written as
\beq{g_eff}
g_{eff}=\frac{GM}{R_A^2}\cos\chi\left(1-\frac{T}{T_{cr}}\right)\,.
\eeq 
The temperature in the quasi-static shell is given by \Eq{e:T(R)}, and 
the condition for the magnetosphere instability can thus be rewritten as:
\beq{m_inst}
\frac{T}{T_{cr}}=\frac{2(\gamma-1)}
{\gamma}\frac{\kappa R_A}{\cos\chi}<1\,.
\eeq 

Consider, for example, the development of the interchange instability when cooling 
(predominantly Compton cooling) is present. The temperature changes as 
\cite{kompaneets56}, \cite{1965PhFl....8.2112W}
\beq{dTdt}
\frac{dT}{dt}=-\frac{T-T_x}{t_C}\,,
\eeq 
where the Compton cooling time is
\beq{t_comp}
t_{C}=\frac{3}{2\mu_m}\frac{\pi R_A^2 m_e c^2}{\sigma_T L_x}
\approx 10.6  [\hbox{s}] R_{9}^2 \dot M_{16}^{-1}\,.
\eeq
Here $m_e$ is the electron mass, $\sigma_T$ is the Thomson cross section, $L_x=0.1 \dot M c^2$ is the X-ray luminosity, $T$ is the electron temperature (which is equal to the ion temperature since the timescale of electron-ion energy exchange here is the shortest possible), $T_x$ is the X-ray temperature and
$\mu_m=0.6$ is the molecular weight. The photon temperature is $T_x=(1/4) T_{cut}$ for a bremsstrahlung spectrum with an exponential cut-off at $T_{cut}$, typically $T_x=3-5$~keV. 
The solution of equation \Eq{dTdt} reads:
\beq{}
T=T_x+(T_{cr}-T_x)e^{-t/t_C}\,.
\eeq 
We note that $T_{cr}\sim 30\,\hbox{keV}\gg T_x\sim 3$~keV. It is seen that for $t\approx 2t_C$ the temperature decreases to $T_x$. In the linear approximation the temperature changes as:
\beq{tlin}
T\approx T_{cr}(1-t/t_C)\,.
\eeq 
Plugging this expression into \Eq{g_eff}, we find that the effective gravity acceleration increases linearly with time as:
\beq{}
g_{eff}\approx \frac{GM}{R_A^2}\frac{t}{t_C}\cos\chi \,.
\eeq 
Correspondingly, the velocity of matter due to the instability growth increases with time as:
\beq{u}
u_i=\int\limits_0^{t} g_{eff} dt=\frac{GM}{R_A^2}\frac{t^2}{2t_C}\cos\chi \,.
\eeq 
Let us introduce the mean rate of the instability growth
\beq{}
<u_i>=\frac{\int u dt}{t}=\frac{1}{6}\frac{GM}{R_A^2}\frac{t^2}{t_C}=
\frac{1}{6}\frac{GM}{R_A^2t_C}\myfrac{\zeta R_A}{<u_i>}^2\cos\chi\,.
\eeq
Here $\zeta\lesssim 1$ and $\zeta R_A$ is the characteristic scale of the
instability that grows with the rate $<u_i>$.
Therefore, for the mean rate of the instability growth at the linear stage we find
\beq{ui}
<u_i>=\myfrac{\zeta^2GM}{6t_C}^{1/3}=\frac{\zeta^{2/3}}{12^{1/3}}\sqrt{\frac{2GM}{R_A}}
\myfrac{t_{ff}}{t_C}^{1/3}\cos\chi\,.
\eeq
As the factor $\cos\chi\simeq 1$, we will omit it below. 
Here we have introduced the characteristic time as
\beq{}
t_{ff}=\frac{R_A^{3/2}}{\sqrt{2GM}}\,,
\eeq
which is close to the free-fall time at a given radius.
Therefore, the factor $f(u)$ becomes:
\beq{fu1}
f(u)=\frac{<u_i>}{u_{ff}(R_A)}\,.
\eeq
Substituting \Eq{ui} and \Eq{fu1} into \Eq{RAdef}, we find for the Alfv\'en radius in 
this regime:
\beq{RAC}
R_A^{(C)}\approx 1.37\times 10^9[\hbox{cm}]
\left(\zeta\frac{\mu_{30}^3}{\dot M_{16}}\right)^{2/11}\,.
\eeq
 Plugging \Eq{RAC} into \Eq{fu1}, we obtain the explicit expression for $f(u)$ in the Compton cooling regime:
\beq{fu}
f(u)_C\approx 0.22
\zeta^{7/11}\dot M_{16}^{4/11}\mu_{30}^{-1/11}\,.
\eeq

In the radiation cooling regime the radiation cooling time is 
\beq{tcrad}
t_{cool}^{(rad)}=\frac{3kT}{2\mu_m n_e \Lambda(T)}=\sqrt{T}/K_{rad},
\eeq
where $\Lambda(T)\approx 2.5\times 10^{-27}\sqrt{T}$ (in CGS units) is the radiation cooling factor (here 
the Gaunt-factor is taken into account and that real cooling function at  high temperatures
goes slightly higher than for pure free-free emission). With this cooling time, temperature
decreases as 
\beq{}
\frac{dT}{dt}=-K_{rad}\sqrt{T}\,,
\eeq
yielding a non-exponential temperature decay with time
\beq{}
\frac{T}{T_0}=\left(1-\frac{1}{2}\frac{K_{rad}t}{\sqrt{T_0}}\right)^2
\eeq 
In the linear approximation, when $t\ll t_{cool}^{(rad)}$, we get for the radiation cooling law
\beq{}
\frac{T}{T_{cr}}=1-\frac{t}{t_{cool}^{(rad)}}\,,
\eeq
similarly to \Eq{tlin} for Compton cooling, and find 
\beq{RArad}
R_A^{(rad)}\approx 1.05\times 10^9[\hbox{cm}]\zeta^{4/81}
\mu_{30}^{16/27}\dot M_{16}^{-6/27}\,,
\eeq
\beq{furad0}
f(u)_{rad}\approx 0.1 \zeta^{14/81}\mu_{30}^{2/27}\dot M_{16}^{6/27}\,.
\eeq

\section{Spin-up/spin-down of neutron star during settling accretion}

At the settling accretion stage onto a NS in a binary system, 
there are three characteristic angular frequencies: 
the angular orbital frequency $\omega_b=2\pi/P_b$, 
which characterizes the specific angular momentum of captured matter, 
the angular frequency of matter near the magnetosphere, $\omega_m(R_A)$, and
the angular frequency of magnetosphere $\omega^*=2\pi/P^*$ which coincides with the NS spin frequency. 
If $\omega_m(R_A)-\omega^*\ne 0$, an effective exchange 
of angular momentum between the magnetosphere and the quasi-spherical shell occurs. 
As shown in Appendices in \cite{2012MNRAS.420..216S}, \cite{2013arXiv1302.0500S},
the rotational law in the shell with settling accretion can be represented in 
a power-law from $\omega(R)\sim 1/R^n$, with $0\le n\le 2$ depending on the
treatment of viscous stresses $W_{R\phi}$ in the shell. In the most likely case where 
anisotropic turbulence appears due to near-sonic convection (see \cite{2012MNRAS.420..216S}), $n\approx 2$, i.e.
iso-angular-momentum rotational law sets in.

The torque due to magnetic forces applied to the neutron star reads:
\beq{torquem}
I\dot \omega^*=\int\frac{B_tB_p}{4\pi}\varpi dS 
\eeq
where $B_t$ is the toroidal magnetic field component which arises if there is the difference of 
the angular velocity of matter $\omega_m$ and magnetosphere angular rotation $\omega^*$.
On the other hand, there is a mechanical torque on the magnetosphere from the base of the shell caused by the turbulent stresses $W_{R\phi}$:
\beq{torquet}
\int W_{R\phi} \varpi dS\,,
\eeq
where the viscous turbulent stresses can be written as 
\beq{torquet1}
W_{R\phi}=\rho \nu_t R \frac{\partial \omega}{\partial R}\,.
\eeq
To specify the turbulent viscosity coefficient
\beq{}
\nu_t=\langle u_c l_t\rangle\,,
\eeq 
we assume that the characteristic scale of the turbulence close to the magnetosphere is 
\beq{lt}
l_t=\zeta_d R_A\,,
\eeq 
where we have introduced the dimensionless factor $\zeta_d\lesssim1$, characterizing the size of the zone in which there is an effective exchange of angular momentum between the magnetosphere and the base of the shell.
The characteristic velocity of the turbulent pulsations $u_c$ 
is determined by the mechanism of turbulence in the plasma above the magnetosphere.
In the case of strong convective motions in the shell, caused by heating of its base, $u_c\sim c_s$, where $c_s$ is the sound speed. 
Equating the torques \Eq{torquem} and \Eq{torquet} and allowing for \Eq{torquet1} and \Eq{lt}, we get
\beq{}
\rho u_c \zeta_d R_A^2 \frac{\partial \omega}{\partial R}=\frac{B_tB_p}{4\pi}
\eeq
We eliminate the density from this expression using the pressure balance at the magnetospheric boundary and the expression for the temperature \Eq{e:T(R)}, and make the substitution 
\beq{zeta}
\frac{\partial \omega}{\partial R}=\frac{\omega_m-\omega^*}{\zeta_d R_A}.
\eeq 

Then we find the relation between the toroidal and poloidal components of the magnetic field in the magnetosphere: 
\beq{btbp}
\frac{B_t}{B_p}=K_2\frac{\gamma}{\sqrt{2}(\gamma-1)}\myfrac{u_c}{u_{ff}}\myfrac{\omega_m-\omega^*}{\omega_K(R_A)}\,.
\eeq
(Note that there is no dependence on the width of the layer characterized by the parameter $\zeta_d$).
Substituting \Eq{btbp} into \Eq{torquem}, the spin-down rate of the neutron star can be written as: 
\beq{sd1}
I\dot\omega^*=K_1K_2\myfrac{u_c}{u_{ff}}\frac{\mu^2}{R_A^3}\frac{\omega_m-\omega^*}{\omega_K(R_A)}\,.
\eeq
where $K_1\sim 1$ is a constant arising from integrating of torques over the magnetospheric surface.

Using the definition of the Alfv\'en radius $R_A$ \Eq{RAdef} and the expression for the Keplerian frequency $\omega_K$, we can write \Eq{sd1} in the form
\beq{sd_om}
I\dot \omega^*=Z \dot M R_A^2(\omega_m-\omega^*).
\eeq
Here the dimensionless coefficient $Z$ is 
\beq{Zdef}
Z=K_1\myfrac{u_c}{u_{ff}}\frac{1}{f(u)}\,.
\eeq

Taking into account that the matter falling onto the neutron star adds the angular momentum
$z\dot M R_A^2\omega^*$, we ultimately get 
\beq{sd_eq}
I\dot \omega^*=Z \dot M R_A^2(\omega_m-\omega^*)+z \dot M R_A^2\omega^*\,.
\eeq
Here $0<z<1$ is the numerical coefficient which is $\sim 2/3$ if matter
enters across the magnetospheric surface with equal probability at different magnetospheric latitudes. 
Substituting $\omega_m(R_A)=\omega_B(R_B/R_A)^2$ for iso-angular-momentum shell, we can rewrite
the above equation in the form
\beq{sd_eq1}
I\dot \omega^*= Z\dot M \omega_B R_B^2-Z(1-z/Z)\dot M R_A^2\omega^*\,.
\eeq 

Substituting for the coupling coefficient $Z$, in the case of Compton cooling we 
can rewrite \Eq{sd_eq}
in the form explicitly showing the spin-up ($K_{su}$) and spin-down ($K_{sd}$) torques:
\beq{sd_eq2}
\dot \omega^*=A\dot M^{\frac{7}{11}} - B\dot M^{3/11}=K_{su}-K_{sd}\,.
\eeq
Here the spin-up/spin-down coefficients $A$ and $B$ do not explicitly depend on $\dot M$.  

For a characteristic value of the accretion rate $\dot M_{16}\equiv \dot M/10^{16}$~g/s, the spin-up and spin-down
torques  read (in CGS units):
\beq{Ksu}
K_{su}\approx 5.29\times 10^{-13}[\hbox{rad/s}^2] K_1\myfrac{u_c}{u_{ff}} \zeta^{-\frac{7}{11}} 
\mu_{30}^{\frac{1}{11}}\myfrac{v_8}{\sqrt{\delta}}^{-4}\myfrac{P_b}{10\hbox{d}}^{-1}\dot M_{16}^{7/11} I_{45}^{-1}
\eeq
\beq{Ksd}
K_{sd}\approx 5.36 \times 10^{-12}[\hbox{rad/s}^2] (1-z/Z) K_1\myfrac{u_c}{u_{ff}} \zeta^{-3/11}\mu_{30}^{{13}/{11}}
\myfrac{P^*}{100\hbox{s}}^{-1}\dot M_{16}^{3/11} I_{45}^{-1}\,.
\eeq
Here $I_{45}=I/10^{45}$~g~cm$^2$ is the NS moment of inertia, 
the dimensionless factor $\delta\sim 1$ takes into account the actual location of the gravitational capture radius.

Another approach to the problem of interaction of quasi-spherically accreting magnetized plasma
with rotating NS magnetospheres is developed in \cite{2014ARep...58..376I}. 

\section{Equilibrium pulsars}

\begin{figure*}
\includegraphics[width=0.45\textwidth]{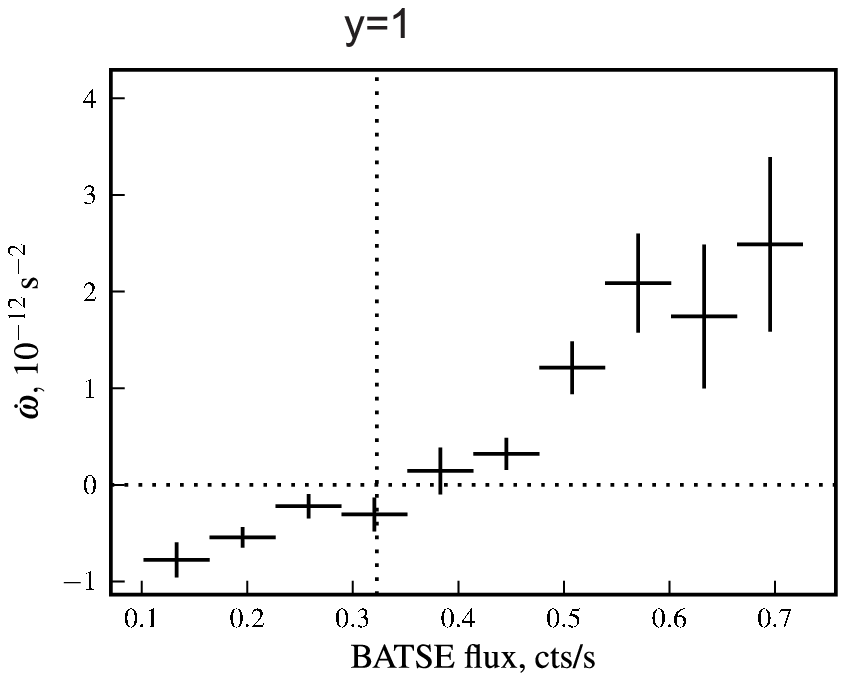}
\includegraphics[width=0.45\textwidth]{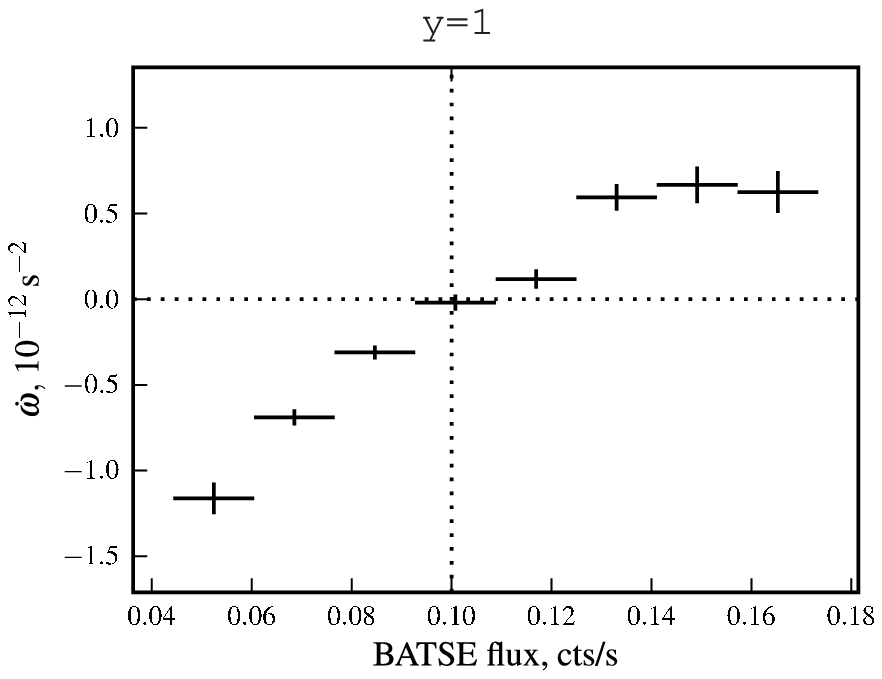}
\parbox[t]{0.47\textwidth}{\caption{Torque-luminosity correlation in GX 301-2, $\dot\omega^*$ as a function of BATSE data (20-40 keV pulsed flux) near the equilibrium frequency \cite{Dor10}. 
The assumed X-ray flux at equilibrium (in terms of
the dimensionless parameter $y$) is also shown by the vertical dotted line.}\label{f:gx}}
\hfill
\parbox[t]{0.47\textwidth}
{\caption{The same as in Fig. \ref{f:gx} for Vela X-1 (V.Doroshenko, PhD Thesis, 2010, IAAT)}\label{f:vela}}
\end{figure*}

For equilibrium pulsars we set 
$\dot \omega^*=0$ and from Equation \Eq{sd_eq} we get
\beq{Zeq}
Z_{eq}(\omega_m-\omega^*)+z\omega^*=0\,.
\eeq 
Close to equilibrium we may vary \Eq{sd_eq} with respect to $\dot M$.
Variations in $\delta \dot M$ may in general be caused by changes in density $\delta \rho$ as well as in velocity of the stellar wind $\delta v$ (and thus the Bondi radius). For density variations only we find (see Eq. (67) in \cite{2013PhyU...56..321S}
for more detail)
\beq{Zeqrho}
Z_{eq,\rho}=\frac{I\frac{\partial \dot\omega^*}{\partial \dot M}|_{eq}}{\frac{4}{11}\omega^*R_A^2}\approx \ 2.52\myfrac{\frac{\partial \dot\omega^*}{\partial y}|_{y=1}}{10^{-12}}\myfrac{P^*}{100s}
\zeta^{-4/11}\dot M_{16}^{-7/11}\mu_{30}^{-12/11}\,.
\eeq
On the other hand, by equating this value to the definition of the coupling coefficient $Z$ (see \Eq{Zdef} above), 
we can find the dimensionless combination of the theory parameters:
\beq{Pi0}
\Pi_0\equiv \frac{K_1\myfrac{u_c}{u_{ff}}}{\zeta^{3/11}}\approx 0.55 \myfrac{\frac{\partial \dot\omega^*}{\partial y}|_{y=1}}{10^{-12}} \myfrac{P^*}{100s}\dot M_{16}^{-3/11}\mu_{30}^{-13/11}\,.
\eeq


The equilibrium period of an X-ray pulsar with known NS magnetic field can be found 
from \Eq{sd_eq1} (or, which is the same, by equating the spin-up and spin-down torques 
from \Eq{Ksu} and \eq{Ksd}):
\beq{Peq}
P_{eq}\approx 1000 [\hbox{s}](1-z/Z_{eq})\zeta^{4/11}\mu_{30,eq}^{12/11}\myfrac{P_b}{10\hbox{d}}
\dot M_{16}^{-4/11}\myfrac{v_8}{\sqrt{\delta}}^{4}\,.
\eeq
In the equilibrium, from this formula we can determine another dimensionless combination of
the theory parameters:
\beq{Pi1}
\Pi_1\equiv \frac{\left(1-\frac{z}{Z_{eq}}\right)\zeta^{4/11}}{\delta^2}\approx 
0.1\myfrac{P^*}{100s}\myfrac{P_b}{10\hbox{d}}^{-1}\dot M_{16}^{4/11}\mu_{30}^{-12/11}v_8^{-4}\,.
\eeq

Because of the strong dependence of the equilibrium period on (usually, poorly measurable) wind velocity, 
for pulsars with independently known magnetic fields $\mu$  
it is more convenient to estimate the wind velocity, assuming $P^*=P^*_{eq}$:
\beq{e:v8min}
v_8\approx 0.56 \left[\frac{\left(1-\frac{z}{Z_{eq}}\right)\zeta^{4/11}}{\delta^2}\right]^{-1/4}\dot M_{16}^{1/11}\mu_{30,eq}^{-3/11}
\myfrac{P_*/100 \hbox{s}}{P_b/10 \hbox{d}}^{1/4}\,,
\eeq
which is only weakly dependent on $\dot M$ and the theory parameter $\Pi_1$. 

In the possible case of mass accretion rate variations due to wind velocity changes only, 
the coupling coefficient $Z_{eq,v}$ reads (see Eq. (68) in \cite{2013PhyU...56..321S}):
\beq{Zeqv}
Z_{eq,v}\approx 0.76 \myfrac{\frac{\partial \dot\omega^*}{\partial y}|_{y=1}}{10^{-12}}\myfrac{P^*}{100s}
\zeta^{-4/11}\dot M_{16}^{-7/11}\mu_{30}^{-12/11}+\frac{7}{10}z\,.
\eeq
Clearly, in this case the coupling is smaller. Below we will consider only 
wind density variations. In principle, if $z>0$ and $(\omega_m-\omega^*)>0$, \Eq{Zeq}
implies that there can be no equilibrium at all -- the pulsar can only spin-up. However, two well-measured
equilibrium pulsars (see below) show that the equilibrium does exist, suggesting that in these objects $(\omega_m-\omega^*)<0$. 

To illustrate the theory outlined above, we show the measured and obtained model parameters of two well-known persistent X-ray puslars, Vela X-1 and GX 301-2 (see Table \ref{t:eq}).

\begin{table*}
\label{t:eq}
 \centering
 \caption{Parameters for the equilibrium X-ray pulsars.} 
$$
\begin{array}{lcc}
\hline
\hbox{Pulsar }&\multicolumn{2}{c}{\hbox{Equilibrium pulsars }} \\
\hline
& {\rm GX 301-2} & {\rm Vela X-1} \\
\hline
\multicolumn{3}{c}{\hbox{Measured parameters}}\\
\hline
P^*{\hbox{(s)}} & 680 & 283 \\
P_B {\hbox{(d)}} & 41.5 & 8.96 \\
v_{w} {\hbox{(km/s)}} & 300? & 700 \\
\mu_{30}& 2.7 & 1.2 \\
\dot M_{16} & 3 & 3 \\
\frac{\partial \dot \omega}{\partial y} \arrowvert_{y=1}{\hbox{(rad/s}^2)} 
& 1.5\cdot10^{-12} & 1.2\cdot10^{-12} \\
\hline
\multicolumn{3}{c}{\hbox{Derived parameters}}\\
\hline
f(u)\zeta^{-7/11} & 0.32 & 0.30 \\
Z_{eq}\zeta^{4/11} & 4.32 & 3.49\\
\Pi_0              & 1.28 & 1.11 \\
v_8\Pi_1^{1/4} \hbox{(km/s)} & 530 & 800\\
\hline
\end{array}
$$
\end{table*}

It is clear from Table \ref{t:eq} that for Vela X-1 observed and derived parameters 
are in good agreement, with the value of dimensionless theory parameters $\Pi_0\sim 1$, as expected
from very general hydrodynamic similarity principles \cite{1959sdmm.book.....S}. It is remarkable that 
parameter $\Pi_0\sim 1$ in GX 301-2 as well, suggesting the common physics of hydrodynamic
interactions in these objects. However, the observed wind velocity in GX 301-2 is inferred
from observations to be around 300 km/s, which is almost two times as small as derived 
from our theory. To obtain such a low velocity from \Eq{e:v8min}, the dimensionless
parameter $\Pi_1$ should be around 10, which is unrealistically high (in fact, this parameter
should not be higher than 1). From this we conclude that in GX 301-2 the wind velocity is likely 
to be estimated not close to the interaction region with NS.

\section{Non-equilibrium pulsars}

\begin{figure*}
\includegraphics[width=0.5\textwidth]{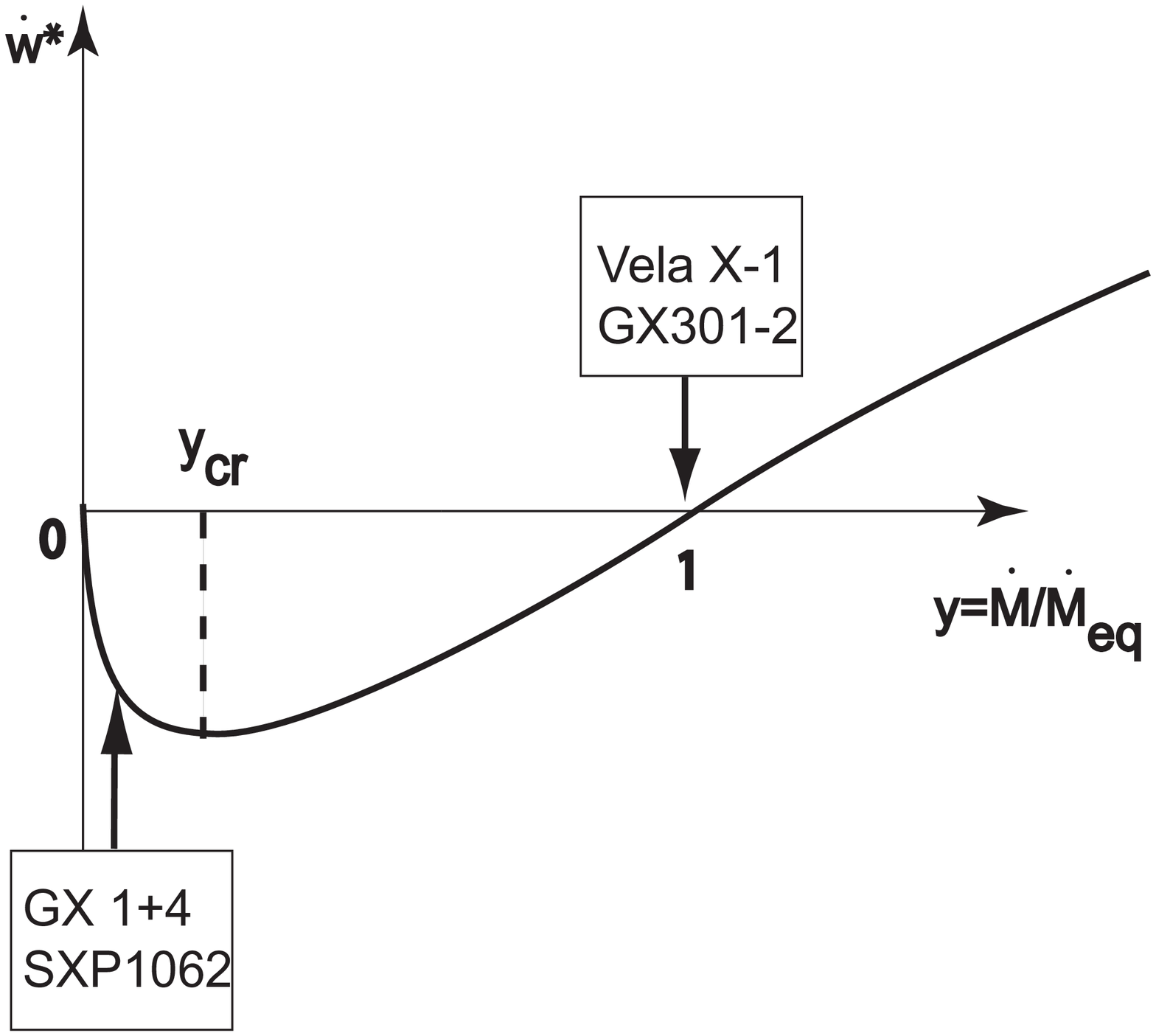}
\caption{Schematics of the dependence of $\dot\omega^*$ on the dimensionless accretion rate $y$. 
 The figure shows the position in the diagram for equilibrium pulsars with $y\sim 1$ and for non-equilibrium pulsars at steady spin-down with $y<y_{cr}$}
\label{f:y}
 \end{figure*}

It is convenient to introduce the dimensionless parameter
\beq{y_def}
y\equiv \frac{\dot M}{\dot M_{eq}}
\eeq
where $\dot M_{eq}$ represents the accretion rate at which $\dot\omega^*=0$:
\beq{dotmeq}
\dot M_{eq}=\myfrac{B}{A}^{11/4}\,.
\eeq
Equation \Eq{sd_eq2} can be rewritten in the form 
\beq{sdy}
I\dot \omega^*=A\dot M_{eq}^\frac{7}{11}y^\frac{7}{11}
\left(1- y^{-\frac{4}{11}}\right)\,.
\eeq
The plot of the function $\dot\omega^*(y)$ is shown schematically in Fig. \ref{f:y}.
The function $\dot\omega^*(\dot M)$ reaches minimum at $\dot M=\dot M_{cr}$:
\beq{}
\dot M_{cr}=\dot M_{eq} \myfrac{3}{7}^\frac{11}{4} \,,
\eeq
In other words, $\dot\omega^*$ attains minimum for the dimensionless parameter
\beq{ycr}
y_{cr}=\myfrac{3}{7}^\frac{11}{4}<1.
\eeq

\begin{table*}
\label{T2}
 \centering
 \caption{} 
 $$
\begin{array}{lccc}
\hline
\hbox{Pulsars }&\multicolumn{3}{c}{\hbox{Non-equilibrium pulsars }}\\
\hline
& {\rm GX 1+4} &{\rm SXP1062}&{\rm 4U 2206+54}\\
\hline
\multicolumn{4}{c}{\hbox{Measured parameters}}\\
\hline
P^*{\hbox{(s)}} & 140 & 1062 &5560\\
P_B {\hbox{(d)}} & 1161 & \sim 300^\dag& 19(?)\\
v_{w} {\hbox{(km/s)}} & 200 & \sim 300^\ddag& 350\\
\mu_{30}&  ? & ? & 1.7\\
\dot M_{16} & 1 & 0.6 & 0.2\\ 
 \dot\omega^*_{sd} &  - 2.34 \cdot 10^{-11} & - 1.63 \cdot 10^{-11} & -9.4 \cdot 10^{-14}\\
\hline
\multicolumn{4}{c}{\hbox{Derived parameters}}\\
\hline
 K_1 (u_c/u_{ff})\zeta^{-3/11} (1-z/Z) & & & 4.3\\
\mu_{30,min}''&\approx 2.4&\approx 10& \approx 0.6 \\
\hline
\end{array}
$$
$^\dag$ Estimate of the source's position in the Corbet diagram
$^\ddag$ Estimate of typical wind vshakuraelocity binary pulsars containing Be-stars.
\end{table*}

The minimum $\dot \omega^*$ for $y=y_{cr}$ (i.e. the maximum possible spin-down rate of the pulsar) is
\beq{omegadotmin}
I\dot \omega^*_{min}=-\frac{4}{3}A\dot M_{eq}^\frac{7}{11}y^\frac{7}{11}\,.
\eeq 

Numerically, the maximum spin-down rate at $y_{cr}$ is 
\beq{e:omegadotsdmax}
\dot\omega^*_{sd,min}\approx -1.12\times 10^{-12}[\hbox{rad/s}^2] (1-z/Z)^{7/4} 
K_1 \myfrac{u_c}{u_{ff}}
\mu_{30}^{2}\myfrac{v_8}{\sqrt{\delta}}^{3}\myfrac{P^*}{100\hbox{s}}^{-7/4}
\myfrac{P_b}{10\hbox{d}}^{3/4}\,.
\eeq
Then, from the condition $|\dot \omega^*_{sd}|\le |\dot\omega^*_{sd,min}|$ 
follows a lower limit on the neutron star magnetic field:
\beq{e:mulim1}
\mu_{30}>\mu_{30, min}'\approx 0.94
\left|\frac{\dot\omega^*_{sd}}{10^{-12}\hbox{rad/s}^2}\right|^{1/2}(1-z/Z)^{-7/8} 
\left(K_1 \myfrac{u_c}{u_{ff}}\right)^{-1/2}
\myfrac{v_8}{\sqrt{\delta}}^{-3/2}\myfrac{P^*}{100\hbox{s}}^{7/8}
\myfrac{P_b}{10\hbox{d}}^{-3/8}.
\eeq

At very small accretion rates $y\ll 1$ the spin-up torque $K_{su}$ can be neglected, and
the spin-down rate of a pulsar is 
\beq{e:sdonly}
\dot \omega^*_{sd}\approx - 0.54\times 10^{-12}[\hbox{rad/s}^2](1-z/Z)K_1 \myfrac{u_c}{u_{ff}}\zeta^{-3/11} 
\mu_{30}^{13/11}\dot M_{16}^{3/11}\myfrac{P^*}{100\hbox{s}}^{-1}.
\eeq
From this we obtain a lower limit on the neutron star magnetic field that does not depend on the stellar wind velocity and the binary orbital period:
\beq{e:mulim2}
\mu_{30}>\mu_{30, min}''\approx 1.68
\left|\frac{\dot\omega^*_{sd}}{10^{-12}\hbox{rad/s}^2}\right|^{11/13} 
(1-z/Z)^{-11/13}\left[K_1 \myfrac{u_c}{u_{ff}} \right]^{-11/13}\zeta^{3/13}\dot M_{16}^{-3/13}
\myfrac{P^*}{100\hbox{s}}^{11/13}\,.
\eeq

As an example, consider the steady spin-down behavior in several slowly rotating moderate-luminosity  
X-ray pulsars (GX 1+4, SXP 1062, shakura4U 2206+54) within the framework of 
quasi-spherical settling accretion theory. The results are summarized in Table 2. 

\section{Bright flares in supergiant fast X-ray transients}

Supergiant Fast X-ray Transients (SFXTs) are a subclass of 
HMXBs 
associated with early-type supergiant companions \cite{Pellizza2006, Chaty2008, Rahoui2008},
and characterized by sporadic, short and bright X--ray flares 
reaching peak luminosities of 10$^{36}$--10$^{37}$~erg~s$^{-1}$.
Most of them were discovered by INTEGRAL \cite{2003ATel..176....1M, 2003ATel..190....1S, 
2003ATel..192....1G, Sguera2005, Negueruela2005}.
They show high dynamic ranges (between 100 and 10,000, depending on the specific source; 
e.g. \cite{Romano2011, 2014A&A...562A...2R}) and their X-ray spectra in outburst are very similar to accreting pulsars in HMXBs.  
In fact, half of them have measured neutron star (NS) spin periods similar to those observed 
from persistent HMXBs (see \cite{2012int..workE..11S} for a review).

The physical mechanism driving their transient behavior, related to the accretion by the compact
object of matter from the supergiant wind, has been discussed by several authors
and is still a matter of debate, as some of them require particular properties of the compact objects hosted in these systems 
\cite{2007AstL...33..149G, 2008ApJ...683.1031B}, 
and others assume
peculiar clumpy properties of the supergiant winds and/or orbital characteristics 
\cite{zand2005,Walter2007,
Sidoli2007,
Negueruela2008,2009MNRAS.398.2152D,Oskinova2012}.

The typical energy released in a SFXT bright flare is about 
$10^{38}-10^{40}$~ergs \cite{2014arXiv1405.5707S}, 
varying by one order of magnitude between different
sources. That is, the mass fallen onto the NS
in a typical bright flare varies from $10^{18}$~g to around $10^{20}$~g. 
shakura
The typical X-ray luminosity outside outbursts in SFXTs is about 
$L_{x,low}\simeq 10^{34}$~erg s$^{-1}$ \cite{2008ApJ...687.1230S},
 and below we shall normalise the luminosity to this value, $L_{34}$. 
At these low X-ray luminosities, the plasma entry rate into the magnetosphere is controlled 
by radiative plasma cooling.
Further, it is convenient to normalise the typical stellar wind velocity from hot OB-supergiants $v_w$ 
to 1000~km~s$^{-1}$ (for orbital periods of about a few days or larger the NS orbital velocities can be neglected compared to the stellar wind velocity from the OB-star), so that the Bondi gravitational capture radius 
is $R_B=2GM/v_w^2=4 \times 10^{10}[\hbox{cm}]v_{8}^{-2}$ 
for a fiducial NS mass of $M_x=1.5 M_\odot$.

\subsection{Magnetopsheric shell instability}

Let us assume that a quasi-statishakurac shell hangs over the magnetosphere around the NS, with the magnetospheric accretion rate being controlled by radiative plasma cooling.
We denote the actual steady-state accretion rate as $\dot M_a$ so that the observed X-ray steady-state luminosity  is $L_x=0.1\dot M_a c^2$. Then from the theory of subsonic 
quasi-spherical accretion \cite{2012MNRAS.420..216S} we know that the factor $f(u)$ (the ratio of the actual velocity of plasma entering the magnetosphere, due to the Rayleigh-Taylor instability, 
to the free-fall velocity at the magnetosphere,
$u_{ff}(R_{A})=\sqrt{2GM/R_A}$) reads \cite{2013MNRAS.428..670S,2014EPJWC..6402001S}
\beq{furad}
f(u)_{rad} \simeq 0.036 \zeta^{7/11} L_{34}^{2/9}\mu_{30}^{2/27}\,.
\eeq
(see also Eq. (\ref{furad0}) above).

The shell is quasi-static (and likely convective).
It is straightforward to calculate the mass of the shell 
using the density distribution $\rho(R)\propto R^{-3/2}$ 
\cite{2012MNRAS.420..216S}. Using the mass continuity equation to eliminate the density above the magnetosphere, we readily find 
\beq{deltaM}
\Delta M \approx \frac{2}{3} \frac{\dot M_a}{f(u)}t_{ff}(R_B)\,.
\eeq
Note that this mass can be expressed through measurable quantities
$L_{x,low}$, $\mu_{30}$ and the (not directly observed) stellar wind
velocity at the Bondi radius $v_w(R_B)$. Using Eq. (\ref{furad}) for  
the radiative plasma cooling, we obtain
\beq{deltaMrad}
\Delta M_{rad} 
\approx 8\times 10^{17} [g] \zeta^{-7/11} L_{34}^{7/9} v_8^{-3}\mu_{30}^{-2/27}\,.
\eeq
The simple estimate (\ref{deltaMrad}) shows that for a typical wind velocity 
near the NS of about 500~km~s$^{-1}$ the \emph{typical} mass of the hot 
magnetospheric shell is around $10^{19}$~g, 
corresponding to $10^{39}$~ergs released in a flare if all the matter from the shell
is accreted onto the NS, as observed. 
Variations in stellar wind velocity between different sources by a factor of $\sim 2$  
would produce the one-order-of-magnitude spread in $\Delta M$ observed in bright SFXT flares.

As noted in \cite{2013MNRAS.428..670S},
if there is an unstable matter flow through the magnetosphere, 
a large quantity of X-ray photons produced near the NS surface should 
rapidly cool down the plasma near the magnetosphere, further increasing the plasma fall velocity
$u_R(R_A)$ 
and the ensuing accretion NS luminosity $L_x$. Therefore, in a bright flare 
the entire shell can fall onto the NS on the free-fall time scale from the outer 
radius of the shell $t_{ff}(R_B)\sim 1000$~s. Clearly, the shell will be replenished by
new wind capture, so the flares will repeat as long as the 
rapid mass entry rate into the magnetosphere is sustained.

\subsection{Magnetized stellar wind as the flare trigger}
\label{sec:magwind}

We suggest that the shell instability described above can be 
triggered by a large-scale magnetic field sporadically 
carried by the stellar wind of the optical OB 
companion. Observations suggest that about $\sim 10\%$ of hot OB-stars have magnetic fields
up to a few kG (see \cite{2013arXiv1312.4755B} for a review and discussion).
It is also well known from Solar wind studies (see e.g. reviews \cite{2004PhyU...47R...1Z, lrsp-2013-2} and references therein) that the Solar wind patches carrying tangent magnetic fields 
has a  lower velocity (about $350$~km~s$^{-1}$) than the wind with radial magnetic fields 
(up to $\sim 700$~km s$^{-1}$). Fluctuations of the stellar wind density and velocity 
from massive stars are also known from spectroscopic observations \cite{2008A&ARv..16..209P}, 
with typical velocity fluctuations up to $0.1\ v_\infty\sim 200-300$~km s$^{-1}$.  

The effect of the magnetic field carried by the stellar wind is twofold: first, 
it may triggershakura
rapid mass entry to the magnetosphere via magnetic reconnection (the phenomenon well known in the dayside Earth magnetosphere, \cite{1961PhRvL...6...47D}), and secondly, 
the magnetized parts of the wind (magnetized clumps with a tangent magnetic field) have a lower velocity than the non magnetized 
ones (or the ones carrying the radial field). As discussed in \cite{2014arXiv1405.5707S} and below,
magnetic reconnection 
can increase the plasma fall velocity in the shell from inefficient, radiative-cooling controlled settling accretion 
with $f(u)_{rad}\sim 0.03-0.1$, 
up to the maximum possible free-fall velocity with $f(u)=1$.
In other words, during a bright flare subsonic 
settling accretion turns into supersonic Bondi accretion.
The second factor (slower wind velocity in magnetized clumps with tangent magnetic field) 
strongly increases the Bondi radius $R_B\propto v_w^{-2}$
and the corresponding Bondi mass accretion rate $\dot M_B\propto v_w^{-3}$. 

Indeed, we can write down the mass accretion rate onto the NS in the unflaring
(low-luminosity) state as $\dot M_{a,low}=f(u) \dot M_B$
with $f(u)$ given by expression (\ref{furad}) 
and $\dot M_B\simeq \pi R_B^2 \rho_w v_w $.
Eliminating the wind density $\rho_w$ using the mass continuity equation written for the 
spherically symmetric stellar wind from the optical star with power $\dot M_o$ and assuming  
a circular binary orbit, we arrive at 
$
\dot M_B\simeq \frac{1}{4}\dot M_o \myfrac{R_B}{a}^2\,.
$
Using the well-known relation for the radiative wind mass-loss rate from massive hot stars
$
\dot M_o\simeq \epsilon \frac{L}{cv_\infty}
$
where $L$ is the optical star luminosity, $v_\infty$ is the stellar wind velocity at infinity,
typically 2000-3000 km s$^{-1}$ for OB stars and $\epsilon\simeq 0.4-1$ is the efficiency factor \cite{1976A&A....49..327L} (in the numerical estimates below we shall assume $\epsilon=0.5$). 
It is also possible to reduce the luminosity $L$ of a
massive star to its mass $M$ using 
the phenomenological relation $(L/L_\odot)\approx 19 (M/M_\odot)^{2.76}$ (see e.g. \cite{2007AstL...33..251V}). Combining the above equations and using
Kepler's third law to express the orbital separation $a$ through the binary period $P_b$, we find for the X-ray luminosity of SFXTs in the non-flaring state 
\begin{eqnarray}
\label{Lxlow}
L_{x,low}\simeq & 5\times 10^{35} [\hbox{erg~s}^{-1}] f(u) 
\myfrac{M}{10 M_\odot}^{2.76-2/3} \nonumber\\
&\myfrac{v_\infty}{1000 \mathrm{km~s}^{-1}}^{-1}
\myfrac{v_w}{500 \mathrm{km~s}^{-1}}^{-4}\myfrac{P_b}{10 \mathrm{d}}^{-4/3}\,,
\end{eqnarray}
which for $f(u)\sim 0.03-0.1$ 
corresponds to the typical low-state luminosities of SFXTs of $\sim 10^{34}$~erg~s$^{-1}$. 

It is straightforward to see that a transition from the low state (subsonic accretion with 
slow magnetospheric entry rate $f(u)\sim 0.03-0.1$) to supersonic free-fall Bondi accretion 
with $f(u)=1$ due to the magnetized stellar wind with the velocity decreasing by a factor of two, for example, would lead to a flaring luminosity of $L_{x,flare}\sim (10\div 30)\times 2^5 L_{x,low}$. This shows that 
the dynamical range of SFXT bright flares ($\sim 300-1000$) can be naturally reproduced by the proposed mechanism.

\subsection{Conditions for magnetic reconnection near the magnetosphere}

For magnetic field reconnection to occur, the time the magnetized plasma spends 
near the magnetopause should be at least comparable to 
the reconnection time, $t_r\sim R_A/v_r$, where
$v_r$ is the magnetic reconnection rate, which is difficult to assess from first principles
\cite{2009ARA&A..47..291Z}.
In real astrophysical plasmas 
the large-scale magnetic reconnection rate can be as high as  
$v_r\sim 0.03-0.07 v_A$ \cite{2009ARA&A..47..291Z}, and phenomenologically 
we can parametrize it as $v_r=\epsilon_r v_A$ 
with $\epsilon_r\sim 0.01-0.1$. The longest time-scale the plasma penetrating into the magnetosphere spends near the magnetopause
is the instability time, $t_{inst}\sim t_{ff}(R_A)f(u)_{rad}$ \cite{2012MNRAS.420..216S}, so the 
reconnection may occur if 
$t_r/t_{inst}\sim (u_{ff}/v_A)(f(u)_{rad}/\epsilon_r)\lesssim 1$. As near 
$R_A$ (from its definition) $v_A\sim u_{ff}$, we arrive at $f(u)_{rad}\lesssim\epsilon_r$ as
the necessary reconnection condition. According to Eq. (\ref{furad}), it is satisfied only 
at sufficiently low X-ray luminosities, pertinent to 'quiet' SFXT states. 
This explains why in HMXBs with convective shells at higher luminosity
(but still lower than $4\times 10^{36}$~erg~s$^{-1}$, at which settling accretion is possible), 
%
reconnection from magnetized plasma accretion will not
lead to the shell instability, but only 
to temporal establishment of the 'strong coupling regime' 
of angular momentum transfer through the shell, as 
discussed in \cite{2012MNRAS.420..216S}.
Episodic strong spin-ups, as observed in GX 301-2, 
may be manifestations of such 'failed' 
reconnection-induced shell instability.

Therefore, it seems likely that the key difference between 
steady HMXBs like Vela X-1, GX 301-2 (showing only moderate flaring activity) and SFXTs is
that in the first case the effects of possibly magnetized stellar winds from optical OB-companions
are insignificant (basically due to the rather high mean accretion rate),
while in SFXTs with lower 'steady' X-ray luminosity, 
large-scale magnetic fields, sporadically carried by clumps in the wind, 
can trigger SFXT flaring activity via magnetic reconnection near the magnetospheric boundary. 
The observed power-law SFXT flare distributions, discussed in \cite{PaizisSidoli2014},
with respect to the log-normal distributions for classical HMXBs \cite{2010A&A...519A..37F}, 
may be related to the properties of magnetized stellar wind and physics of its interaction 
with the NS magnetosphere \cite{2014MNRAS.442.2325S,2016MNRAS.457.3693S}.
14-12-00146
\vskip\baselineskip

\textbf{Acknowledgement.} The work is supported by the Russian Science Foundation grant 14-12-00146.

\bibliographystyle{azh}
\bibliography{wind_accretion}

\end{document}